\documentclass[14pt]{revtex4}
\usepackage{amssymb}
\usepackage{latexsym}
\usepackage{epsfig}
\begin{document}                             

\title{Oscillation of cosmic space in the background of two interacting tachyonic  BIons}

\author{ Aroonkumar Beesham$^{1}$\footnote{beeshama@unizulu.ac.za}}
\affiliation{$^{1}$Department of Mathematical Sciences, University of Zululand, Private Bag X1001, Kwa-Dlangezwa 3886, South Africa}
\affiliation{$^{2}$Faculty of Natural Sciences, Mangosuthu University of Technology, P O Box 12363, Jacobs 4026, South Africa}

\begin{abstract}
When a brane-anti-brane system includes two tachyons, each of them produces a BIoninc wormhole. These wormholes interact with each other and form 4 regions. Two of the regions are related to the independent BIons which have been considered previously. However, two new regions  correspond to the interacting BIons in  which the cosmic parameters act oppositely to each other.  We obtain the Hubble parameter and energy density of the universes in the new regions and show that by expanding a universe in one region, the universe in the other region contracts. Also, the evolution of the universes depend on the tachyonic fields, the separation between the branes and the size of the throats of the bionic wormholes.
\end{abstract}

\maketitle

\noindent
{\it keywords:} Accelerating BIon; Padmanabhan mechanism; Extra dimensions; Birth of universe; Expansion of universe; Degrees of freedom

\section{Introduction}
Recently,  Padmanabhan has proposed a mechanism for the formation of cosmic space and proposed that the accelerated expansion of the universe could emerge from  the difference
between the surface degrees of freedom on the holographic horizon and the bulk degrees of freedom  \cite{q1}. Following this mechanism , many discussions have been done by scientists \cite{q2,q3,q4,q5,q6,q7,q8}. For example, in one research, using this mechanism, the Friedmann equations of an (n + 1)-dimensional
Friedmann-Lemaitre-Robertson-Walker universe in the back ground of  Gauss-Bonnet gravity and Lovelock gravity have been formulated \cite{q2}. Another article has used the idea of treating the cosmic space as an emergent process in brane cosmology, scalar-tensor cosmology, and f(R) gravity, and  has obtained the corresponding
cosmological equations in these theories \cite{q3}. In another work, applying the Padmanabhan mechanism, investigators have obtained the Friedmann equations of the universe not only in four dimensional space-time, but also in higher dimensional space-time and  other gravity theories \cite{q4}. In other investigation, by considering the invariant volume surrounded by the apparent horizon, the evolution
equation in the Padmanabhan idea has been generalized to give the Friedmann equation  in the nonflat universe related to
$k = \pm 1$  \cite{q5}. In another work, the author used  the equations of the universe calculated  in the Padmanabhan model  with the corrected
entropy-area law that follows from the Generalized Uncertainty Principle (GUP) and obtained a modified
Friedmann equation due to the GUP \cite{q6}. Other authors have constructed the Padmanabhan idea  in a BIonic system and  shown that all degrees of freedom inside and outside the universe are controlled by the evolution of the BIon in the extra dimensions \cite{q7}. And finally, in a latest article, the emergence and expansion of cosmic space in the background of an accelerating BIon has been considered \cite{q8}. Motivated by these works, we construct the Padmanabhan model in the background of two interacting tachyonic BIonic wormholes. A BIon is a system of two branes which are connected by wormholes. We will show that the interaction of wormholes leads to the emergence of two accelerations and four regions. We will obtain the cosmic parameters in each region in terms of tachtonic potentials.
 
 The outline of the
paper is as  follows. In section II, we consider the background of two interacting BIons. In section III, we construct the cosmic space in this background and obtain the Hubble parameters in terms of the BIonic parameters. The last section is devoted to the conclusion.

\section{ Two interacting tachyonic  BIons}\label{o1}

In this section, we consider a system of two branes which are connected by two interacting wormholes. In this system, a tachyonic field causes the acceleration and emergence of two regions. Then, the second tachyon changes the background, and produce two new regions (See figures 1 and 2).  
\begin{figure*}[thbp]
	\begin{center}
		\begin{tabular}{rl}
			\includegraphics[width=8cm]{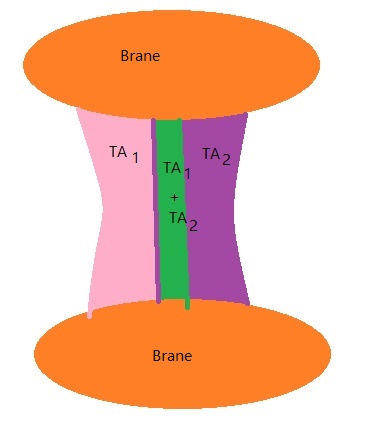}
		\end{tabular}
	\end{center}
	\caption{Two interacting tachyonic BIonic wormholes}
\end{figure*}

\begin{figure*}[thbp]
	\begin{center}
		\begin{tabular}{rl}
			\includegraphics[width=8cm]{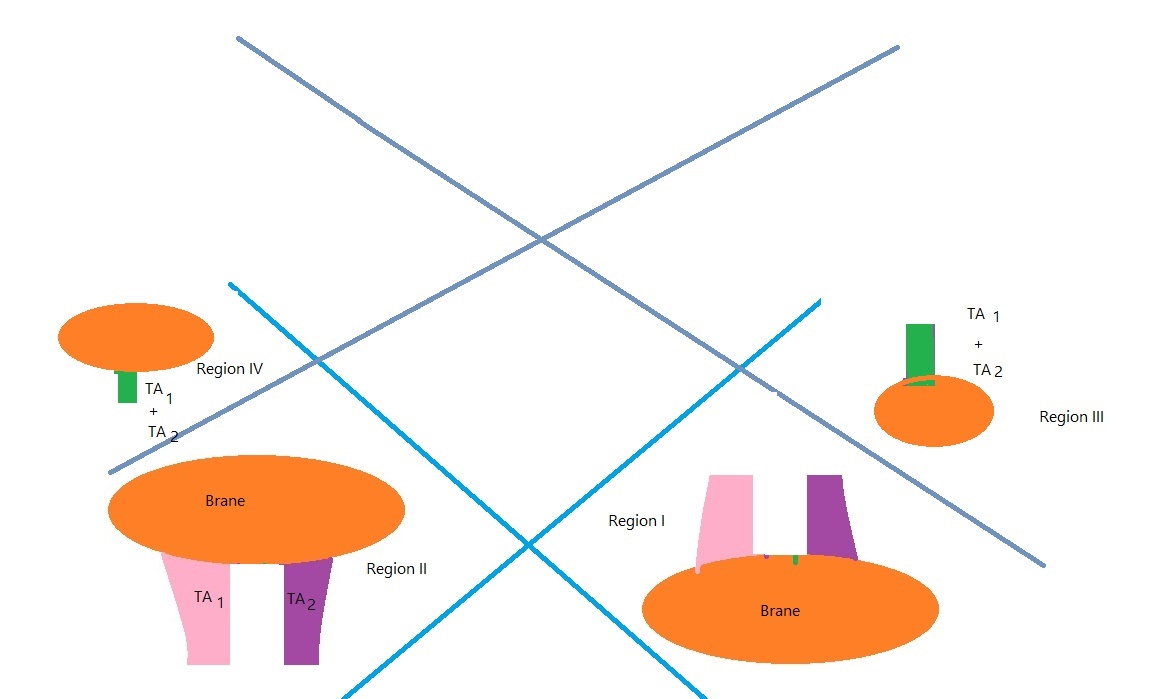}
		\end{tabular}
	\end{center}
	\caption{Emergence of 4 regions by tachyonic  BIonic wormholes}
\end{figure*}

Previously, it has been shown that the  metric of a thermal BIon in 10-dimensional space-time is given by \cite{q8, q9,q10}

\begin{eqnarray}
&& ds^{2} = D^{-\frac{1}{2}} H^{-\frac{1}{2}}\Big(dx_{2}^{2} + dx_{3}^{2}\Big) + D^{\frac{1}{2}} H^{-\frac{1}{2}}\Big(-f dt^{2} + dx_{1}^{2}\Big) + D^{-\frac{1}{2}} H^{\frac{1}{2}}\Big(f^{-1} dr^{2} + r^{2}d\Omega_{5}^{2}\Big)\nonumber\\&& 
\label{a1}
\end{eqnarray} 

where

\begin{eqnarray}
&& f = 1-\frac{r_{0}^{4}}{r^{4}} \quad H = 1+\frac{r_{0}^{4}\sinh^{2}\alpha}{r^{4}}\quad D = \cos^{2}\epsilon + \sin^{2}\epsilon H^{-1}
\label{a2}
\end{eqnarray}

and

\begin{eqnarray}
&& \cosh^{2} \alpha = \frac{3}{2}\frac{\cos\frac{\delta}{3} + \sqrt{3}\cos\frac{\delta}{3}}{\cos\delta}\quad \cos\epsilon = \frac{1}{\sqrt{1 + \frac{K^{2}}{r^{4}}}}
\label{a3}
\end{eqnarray}

The angle $\delta$ is defined as:

\begin{eqnarray}
&& \cos\delta = \bar{T}^{4}\sqrt{1 + \frac{K^{2}}{r^{4}}}\quad   \bar{T} = \Big(\frac{9\pi^{2}N}{4\sqrt{3}T_{D3}}\Big)^{\frac{1}{2}}T
\label{a4}
\end{eqnarray}
 
 A tachyon creates a potential. This potential produces a force and an acceleration. This acceleration changes the background of the system. For an accelerating tachyonic BIonic wormhole, the relation between the world volume of the coordinates of the accelerating D3-branes ($\tau_{1}, \sigma_{1} $) and the coordinates of 10D
Minkowski space-time ($t_{1}, r_{1}$) are \cite{q11}:

\begin{eqnarray}
&& a_{1}t_{1}= e^{a_{1}\sigma_{1}} \sinh(a_{1}\tau_{1}) \quad a_{1}r_{1}=e^{a_{1}\sigma_{1}} \cosh(a_{1}\tau_{1}) \quad \text{In Region I} \nonumber\\&& a_{1}t_{1}= - e^{-a_{1}\sigma_{1}} \sinh(a_{1}\tau_{1}) \quad a_{1}r_{1} =  e^{-a_{1}\sigma_{1}} \cosh(a_{1}\tau_{1}) \quad \text{In Region II}
\label{a5}
\end{eqnarray}
where $a$ is the acceleration of the BIon. Now, we put this system in the background of the other accelerating tachyonic wormhole. We can write:

\begin{eqnarray}
&& a_{2}t_{2}= e^{a_{2}r_{1}} \sinh(a_{2}t_{1}) \quad a_{2}r_{2}=e^{a_{2}r_{1}} \cosh(a_{2}t_{1}) \quad \text{In Region III} \nonumber\\&& a_{2}t_{2}= - e^{-a_{2}r_{1}} \sinh(a_{2}t_{1}) \quad a_{2}r_{2} =  e^{-a_{2}r_{1}} \cosh(a_{2}t_{1}) \quad \text{In Region IV}
\label{a6}
\end{eqnarray}

or equivalently:

\begin{eqnarray}
&&  a_{2}t_{2}= e^{\frac{a_{2}}{a_{1}}e^{a_{1}\sigma_{1}} \cosh(a_{1}\tau_{1})} \sinh(\frac{a_{2}}{a_{1}}e^{a_{1}\sigma_{1}} \sinh(a_{1}\tau_{1}))\nonumber\\&& a_{2}r_{2}=e^{\frac{a_{2}}{a_{1}}e^{a_{1}\sigma_{1}} \cosh(a_{1}\tau_{1})} \cosh(\frac{a_{2}}{a_{1}}e^{a_{1}\sigma_{1}} \sinh(a_{1}\tau_{1})) \nonumber\\&&\quad \quad \quad\quad\quad \quad \quad \quad \quad \quad \quad\quad \quad \quad \quad \quad\text{In Region III} \nonumber\\&& a_{2}t_{2}= - e^{-\frac{a_{2}}{a_{1}}e^{-a_{1}\sigma_{1}} \cosh(a_{1}\tau_{1})} \sinh(-\frac{a_{2}}{a_{1}}e^{-a_{1}\sigma_{1}} \sinh(a_{1}\tau_{1})) \nonumber\\&& a_{2}r_{2} =  e^{-\frac{a_{2}}{a_{1}}e^{-a_{1}\sigma_{1}} \cosh(a_{1}\tau_{1})} \cosh(-\frac{a_{2}}{a_{1}}e^{-a_{1}\sigma_{1}} \sinh(a_{1}\tau_{1}))\nonumber\\&& \quad \quad \quad \quad\quad\quad \quad \quad \quad\quad \quad \quad \quad \quad \quad \quad \text{In Region IV}
\label{a7}
\end{eqnarray}

These accelerations have a direct relation with the effective potentials in each region:

\begin{eqnarray}
&&  a_{1} \sim \frac{\partial V_{1}}{m_{1}\partial \sigma_{1}}   \nonumber\\&& a_{2} \sim \frac{\partial V_{2}}{m_{2}\partial \sigma_{2}} 
\label{a8}
\end{eqnarray}

Substituting the accelerations from equation (\ref{a8}) in equation (\ref{a7}), we obtain:

\begin{eqnarray}
&& \frac{\partial V_{2}}{m_{2}\partial \sigma_{2}} t_{2}= e^{\frac{\frac{\partial V_{2}}{m_{2}\partial \sigma_{2}} }{\frac{\partial V_{2}}{m_{2}\partial \sigma_{2}} }e^{\frac{\partial V_{2}}{m_{2}\partial \sigma_{2}} \sigma_{1}} \cosh(a_{1}\tau_{1})} \sinh(\frac{\frac{\partial V_{2}}{m_{2}\partial \sigma_{2}} }{\frac{\partial V_{2}}{m_{2}\partial \sigma_{2}} }e^{\frac{\partial V_{2}}{m_{2}\partial \sigma_{2}} \sigma_{1}} \sinh(\frac{\partial V_{2}}{m_{2}\partial \sigma_{2}} \tau_{1}))\nonumber\\&& \frac{\partial V_{2}}{m_{2}\partial \sigma_{2}} r_{2}=e^{\frac{\frac{\partial V_{2}}{m_{2}\partial \sigma_{2}} }{\frac{\partial V_{2}}{m_{2}\partial \sigma_{2}} }e^{\frac{\partial V_{2}}{m_{2}\partial \sigma_{2}} \sigma_{1}} \cosh(\frac{\partial V_{2}}{m_{2}\partial \sigma_{2}} \tau_{1})} \cosh(\frac{\frac{\partial V_{2}}{m_{2}\partial \sigma_{2}} }{\frac{\partial V_{2}}{m_{2}\partial \sigma_{2}} }e^{\frac{\partial V_{2}}{m_{2}\partial \sigma_{2}} \sigma_{1}} \sinh(a_{1}\tau_{1})) \nonumber\\&&\quad \quad \quad\quad\quad \quad \quad \quad \quad \quad \quad  \quad \quad \quad \quad \quad \quad\quad \quad \quad \quad \quad\text{In Region III} \nonumber\\&& \frac{\partial V_{2}}{m_{2}\partial \sigma_{2}} t_{2}= - e^{-\frac{\frac{\partial V_{2}}{m_{2}\partial \sigma_{2}} }{\frac{\partial V_{2}}{m_{2}\partial \sigma_{2}} }e^{-\frac{\partial V_{2}}{m_{2}\partial \sigma_{2}} \sigma_{1}} \cosh(\frac{\partial V_{2}}{m_{2}\partial \sigma_{2}} \tau_{1})} \sinh(-\frac{\frac{\partial V_{2}}{m_{2}\partial \sigma_{2}} }{\frac{\partial V_{2}}{m_{2}\partial \sigma_{2}} }e^{-\frac{\partial V_{2}}{m_{2}\partial \sigma_{2}} \sigma_{1}} \sinh(\frac{\partial V_{2}}{m_{2}\partial \sigma_{2}} \tau_{1})) \nonumber\\&& \frac{\partial V_{2}}{m_{2}\partial \sigma_{2}} r_{2} =  e^{-\frac{\frac{\partial V_{2}}{m_{2}\partial \sigma_{2}} }{\frac{\partial V_{2}}{m_{2}\partial \sigma_{2}} }e^{-\frac{\partial V_{2}}{m_{2}\partial \sigma_{2}} \sigma_{1}} \cosh(a_{1}\tau_{1})} \cosh(-\frac{\frac{\partial V_{2}}{m_{2}\partial \sigma_{2}} }{\frac{\partial V_{2}}{m_{2}\partial \sigma_{2}} }e^{-\frac{\partial V_{2}}{m_{2}\partial \sigma_{2}} \sigma_{1}} \sinh(\frac{\partial V_{2}}{m_{2}\partial \sigma_{2}} \tau_{1}))\nonumber\\&& \quad \quad \quad \quad\quad\quad \quad \quad \quad\quad \quad \quad \quad \quad \quad \quad \quad \quad \quad \quad \quad \quad \text{In Region IV}
\label{a9}
\end{eqnarray}

For the potentials $V_{1/2}$, we make use of the tachyonic Hamiltonians: \cite{q12,q13,q14}:
\begin{eqnarray}
&& H_{tachyonic} = T + V_{tot} \nonumber \\&& V_{tot}\simeq  \int d \sigma V(TA)\sqrt{D_{TA}}F_{DBI},  \nonumber \\&&
F_{DBI}=\sigma^{2}\sqrt{1 + \frac{\beta}{\sigma^{4}}}\nonumber \\&& D_{TA} = 1 + \frac{l'(\sigma)^{2}}{4}+ TA^{2}l^{2},\nonumber \\&& V(TA)=\frac{\zeta_{3}}{\cosh\sqrt{\pi}TA}\label{a10}
\end{eqnarray}

where $TA$ is the tachyonic field and $l$ is the separation between the two branes in a BIon or the length of a BIon. Also, derivatives ($l'$) are with respect to $\sigma$. Substituting equations (\ref{a10}) into equation (\ref{a8}), we obtain:

\begin{eqnarray}
&&  a_{1} \sim \frac{V(TA_{1})\sqrt{1 + \frac{l'(\sigma_{1})^{2}}{4}+ TA_{1}^{2}l_{1}^{2}}\sigma_{1}^{2}\sqrt{1 + \frac{\beta}{\sigma_{1}^{4}}}}{m_{TA_{1}}}   \nonumber\\&& a_{2} \sim  \frac{V(TA_{2})\sqrt{1 + \frac{l'(\sigma_{2})^{2}}{4}+ TA_{2}^{2}l_{2}^{2}}\sigma_{2}^{2}\sqrt{1 + \frac{\beta}{\sigma_{2}^{4}}}}{m_{TA_{2}}}
\label{a11}
\end{eqnarray}

The above equation shows that the accelerations depend on the tachyonic potentials, separation between the branes and the tachyonic masses. This means that the tachyonic fields control the evolutions of the  universe. Substituting equation (\ref{a11}) into equation (\ref{a7}), we get:

\begin{eqnarray}
&&  a_{2}t_{2}= F(TA_{1},TA_{2},\sigma_{1},\sigma_{2})=\nonumber\\&& e^{\frac{\frac{V(TA_{2})\sqrt{1 + \frac{l'(\sigma_{2})^{2}}{4}+ TA_{2}^{2}l_{2}^{2}}\sigma_{2}^{2}\sqrt{1 + \frac{\beta}{\sigma_{2}^{4}}}}{m_{TA_{2}}}}{\frac{V(TA_{1})\sqrt{1 + \frac{l'(\sigma_{1})^{2}}{4}+ TA_{1}^{2}l_{1}^{2}}\sigma_{1}^{2}\sqrt{1 + \frac{\beta}{\sigma_{1}^{4}}}}{m_{TA_{1}}}}e^{\frac{V(TA_{1})\sqrt{1 + \frac{l'(\sigma_{1})^{2}}{4}+ TA_{1}^{2}l_{1}^{2}}\sigma_{1}^{2}\sqrt{1 + \frac{\beta}{\sigma_{1}^{4}}}}{m_{TA_{1}}}\sigma_{1}} \cosh(\frac{V(TA_{1})\sqrt{1 + \frac{l'(\sigma_{1})^{2}}{4}+ TA_{1}^{2}l_{1}^{2}}\sigma_{1}^{2}\sqrt{1 + \frac{\beta}{\sigma_{1}^{4}}}}{m_{TA_{1}}}\tau_{1})} \times \nonumber\\&& \sinh(\frac{\frac{V(TA_{2})\sqrt{1 + \frac{l'(\sigma_{2})^{2}}{4}+ TA_{2}^{2}l_{2}^{2}}\sigma_{2}^{2}\sqrt{1 + \frac{\beta}{\sigma_{2}^{4}}}}{m_{TA_{2}}}}{\frac{V(TA_{1})\sqrt{1 + \frac{l'(\sigma_{1})^{2}}{4}+ TA_{1}^{2}l_{1}^{2}}\sigma_{1}^{2}\sqrt{1 + \frac{\beta}{\sigma_{1}^{4}}}}{m_{TA_{1}}}}e^{\frac{V(TA_{1})\sqrt{1 + \frac{l'(\sigma_{1})^{2}}{4}+ TA_{1}^{2}l_{1}^{2}}\sigma_{1}^{2}\sqrt{1 + \frac{\beta}{\sigma_{1}^{4}}}}{m_{TA_{1}}}\sigma_{1}} \times \nonumber\\&&\sinh(\frac{V(TA_{1})\sqrt{1 + \frac{l'(\sigma_{1})^{2}}{4}+ TA_{1}^{2}l_{1}^{2}}\sigma_{1}^{2}\sqrt{1 + \frac{\beta}{\sigma_{1}^{4}}}}{m_{TA_{1}}}\tau_{1}))\nonumber\\&&\nonumber\\&&\nonumber\\&& a_{2}r_{2}=E(TA_{1},TA_{2},\sigma_{1},\sigma_{2})=\nonumber\\&& e^{\frac{\frac{V(TA_{2})\sqrt{1 + \frac{l'(\sigma_{2})^{2}}{4}+ TA_{2}^{2}l_{2}^{2}}\sigma_{2}^{2}\sqrt{1 + \frac{\beta}{\sigma_{2}^{4}}}}{m_{TA_{2}}}}{\frac{V(TA_{1})\sqrt{1 + \frac{l'(\sigma_{1})^{2}}{4}+ TA_{1}^{2}l_{1}^{2}}\sigma_{1}^{2}\sqrt{1 + \frac{\beta}{\sigma_{1}^{4}}}}{m_{TA_{1}}}}e^{\frac{V(TA_{1})\sqrt{1 + \frac{l'(\sigma_{1})^{2}}{4}+ TA_{1}^{2}l_{1}^{2}}\sigma_{1}^{2}\sqrt{1 + \frac{\beta}{\sigma_{1}^{4}}}}{m_{TA_{1}}}\sigma_{1}} \cosh(\frac{V(TA_{1})\sqrt{1 + \frac{l'(\sigma_{1})^{2}}{4}+ TA_{1}^{2}l_{1}^{2}}\sigma_{1}^{2}\sqrt{1 + \frac{\beta}{\sigma_{1}^{4}}}}{m_{TA_{1}}}\tau_{1})} \times \nonumber\\&&\cosh(\frac{\frac{V(TA_{2})\sqrt{1 + \frac{l'(\sigma_{2})^{2}}{4}+ TA_{2}^{2}l_{2}^{2}}\sigma_{2}^{2}\sqrt{1 + \frac{\beta}{\sigma_{2}^{4}}}}{m_{TA_{2}}}}{\frac{V(TA_{1})\sqrt{1 + \frac{l'(\sigma_{1})^{2}}{4}+ TA_{1}^{2}l_{1}^{2}}\sigma_{1}^{2}\sqrt{1 + \frac{\beta}{\sigma_{1}^{4}}}}{m_{TA_{1}}}}e^{\frac{V(TA_{1})\sqrt{1 + \frac{l'(\sigma_{1})^{2}}{4}+ TA_{1}^{2}l_{1}^{2}}\sigma_{1}^{2}\sqrt{1 + \frac{\beta}{\sigma_{1}^{4}}}}{m_{TA_{1}}}\sigma_{1}}\times \nonumber\\&& \sinh(\frac{V(TA_{1})\sqrt{1 + \frac{l'(\sigma_{1})^{2}}{4}+ TA_{1}^{2}l_{1}^{2}}\sigma_{1}^{2}\sqrt{1 + \frac{\beta}{\sigma_{1}^{4}}}}{m_{TA_{1}}}\tau_{1})) \nonumber\\&&\quad \quad \quad\quad\quad \quad \quad \quad \quad \quad \quad\quad \quad \quad \quad \quad\text{In Region III} \nonumber\\&& \nonumber\\&&\nonumber\\&&\nonumber\\&&a_{2}t_{2}= G(TA_{1},TA_{2},\sigma_{1},\sigma_{2}) \nonumber\\&&- e^{-\frac{\frac{V(TA_{2})\sqrt{1 + \frac{l'(\sigma_{2})^{2}}{4}+ TA_{2}^{2}l_{2}^{2}}\sigma_{2}^{2}\sqrt{1 + \frac{\beta}{\sigma_{2}^{4}}}}{m_{TA_{2}}}}{\frac{V(TA_{1})\sqrt{1 + \frac{l'(\sigma_{1})^{2}}{4}+ TA_{1}^{2}l_{1}^{2}}\sigma_{1}^{2}\sqrt{1 + \frac{\beta}{\sigma_{1}^{4}}}}{m_{TA_{1}}}}e^{-a_{1}\sigma_{1}} \cosh(a_{1}\tau_{1})} \times \nonumber\\&&\sinh(-\frac{\frac{V(TA_{2})\sqrt{1 + \frac{l'(\sigma_{2})^{2}}{4}+ TA_{2}^{2}l_{2}^{2}}\sigma_{2}^{2}\sqrt{1 + \frac{\beta}{\sigma_{2}^{4}}}}{m_{TA_{2}}}}{\frac{V(TA_{1})\sqrt{1 + \frac{l'(\sigma_{1})^{2}}{4}+ TA_{1}^{2}l_{1}^{2}}\sigma_{1}^{2}\sqrt{1 + \frac{\beta}{\sigma_{1}^{4}}}}{m_{TA_{1}}}}e^{-\frac{V(TA_{1})\sqrt{1 + \frac{l'(\sigma_{1})^{2}}{4}+ TA_{1}^{2}l_{1}^{2}}\sigma_{1}^{2}\sqrt{1 + \frac{\beta}{\sigma_{1}^{4}}}}{m_{TA_{1}}}\sigma_{1}}\times \nonumber\\&& \sinh(\frac{V(TA_{1})\sqrt{1 + \frac{l'(\sigma_{1})^{2}}{4}+ TA_{1}^{2}l_{1}^{2}}\sigma_{1}^{2}\sqrt{1 + \frac{\beta}{\sigma_{1}^{4}}}}{m_{TA_{1}}}\tau_{1})) \nonumber\\&& \nonumber\\&& \nonumber\\&&\nonumber\\&&\nonumber\\&& a_{2}r_{2} = O (TA_{1},TA_{2},\sigma_{1},\sigma_{2})=\nonumber\\&& e^{-\frac{\frac{V(TA_{2})\sqrt{1 + \frac{l'(\sigma_{2})^{2}}{4}+ TA_{2}^{2}l_{2}^{2}}\sigma_{2}^{2}\sqrt{1 + \frac{\beta}{\sigma_{2}^{4}}}}{m_{TA_{2}}}}{\frac{V(TA_{1})\sqrt{1 + \frac{l'(\sigma_{1})^{2}}{4}+ TA_{1}^{2}l_{1}^{2}}\sigma_{1}^{2}\sqrt{1 + \frac{\beta}{\sigma_{1}^{4}}}}{m_{TA_{1}}}}e^{-\frac{V(TA_{1})\sqrt{1 + \frac{l'(\sigma_{1})^{2}}{4}+ TA_{1}^{2}l_{1}^{2}}\sigma_{1}^{2}\sqrt{1 + \frac{\beta}{\sigma_{1}^{4}}}}{m_{TA_{1}}}\sigma_{1}} \cosh(\frac{V(TA_{1})\sqrt{1 + \frac{l'(\sigma_{1})^{2}}{4}+ TA_{1}^{2}l_{1}^{2}}\sigma_{1}^{2}\sqrt{1 + \frac{\beta}{\sigma_{1}^{4}}}}{m_{TA_{1}}}\tau_{1})} \times \nonumber\\&&\cosh(-\frac{\frac{V(TA_{2})\sqrt{1 + \frac{l'(\sigma_{2})^{2}}{4}+ TA_{2}^{2}l_{2}^{2}}\sigma_{2}^{2}\sqrt{1 + \frac{\beta}{\sigma_{2}^{4}}}}{m_{TA_{2}}}}{\frac{V(TA_{1})\sqrt{1 + \frac{l'(\sigma_{1})^{2}}{4}+ TA_{1}^{2}l_{1}^{2}}\sigma_{1}^{2}\sqrt{1 + \frac{\beta}{\sigma_{1}^{4}}}}{m_{TA_{1}}}}e^{-\frac{V(TA_{1})\sqrt{1 + \frac{l'(\sigma_{1})^{2}}{4}+ TA_{1}^{2}l_{1}^{2}}\sigma_{1}^{2}\sqrt{1 + \frac{\beta}{\sigma_{1}^{4}}}}{m_{TA_{1}}}\sigma_{1}} \times \nonumber\\&&\sinh(\frac{V(TA_{1})\sqrt{1 + \frac{l'(\sigma_{1})^{2}}{4}+ TA_{1}^{2}l_{1}^{2}}\sigma_{1}^{2}\sqrt{1 + \frac{\beta}{\sigma_{1}^{4}}}}{m_{TA_{1}}}\tau_{1}))\nonumber\\&& \quad \quad \quad \quad\quad\quad \quad \quad \quad\quad \quad \quad \quad \quad \quad \quad \text{In Region IV}
\label{a12}
\end{eqnarray}

The above equation shows how the interaction between the two tachyonic wormholes changes the background and the metric of the system. Substituting the results of equation (\ref{a12}) into equation (\ref{a1}), we obtain :

\begin{eqnarray}
&& ds^{2}_{III,A,thermal}= D_{III-A}^{\frac{1}{2}} H_{III-A}^{-\frac{1}{2}}f_{III-A}\Big(e^{2 E(TA_{1},TA_{2},\sigma_{1},\sigma_{2})}\nonumber\\&&  + \frac{1}{\sinh^{2}(F(TA_{1},TA_{2},\sigma_{1},\sigma_{2}))}(\frac{dz}{d t_{2}})^{2} \Big)dt_{2}^{2} -  \nonumber\\&& D_{III-A}^{-\frac{1}{2}} H_{III-A}^{\frac{1}{2}}f_{III-A}^{-1}\Big(e^{2E(TA_{1},TA_{2},\sigma_{1},\sigma_{2})}+ \nonumber\\&& \frac{1}{\cosh^{2}(F(TA_{1},TA_{2},\sigma_{1},\sigma_{2}))}(\frac{dz}{d r_{2}})^{2} \Big) d r_{2}^{2} + \nonumber\\&& \frac{1}{\sinh(F(TA_{1},TA_{2},\sigma_{1},\sigma_{2}))\cosh(F(TA_{1},TA_{2},\sigma_{1},\sigma_{2}))}(\frac{dz}{dt_{2} }\frac{dz}{dr_{2}})dt_{2} dr_{2} +  \nonumber\\&&  D_{III-A}^{-\frac{1}{2}} H_{III-A}^{\frac{1}{2}}\Big(\frac{1}{a}e^{E(TA_{1},TA_{2},\sigma_{1},\sigma_{2})} \cosh(F(TA_{1},TA_{2},\sigma_{1},\sigma_{2}))\Big)^{2}\Big(d\theta^{2} + sin^{2}\theta d\phi^{2}\Big)  + \nonumber\\&& D_{III-A}^{-\frac{1}{2}} H_{III-A}^{-\frac{1}{2}} \sum_{i=1}^{5}dx_{i}^{2}
\label{a13}
\end{eqnarray}

\begin{eqnarray}
&& ds^{2}_{IV,A,thermal}= D_{IV-A}^{\frac{1}{2}} H_{IV-A}^{-\frac{1}{2}}f_{IV-A}\Big(e^{-2O(TA_{1},TA_{2},\sigma_{1},\sigma_{2})} +\nonumber\\&&  \frac{1}{\sinh^{2}(G(TA_{1},TA_{2},\sigma_{1},\sigma_{2}))}(\frac{dz}{dt_{2}})^{2} \Big)dt_{2}^{2} - \nonumber\\&& D_{IV-A}^{-\frac{1}{2}} H_{IV-A}^{\frac{1}{2}}f_{IV-A}^{-1} \Big(e^{-2O(TA_{1},TA_{2},\sigma_{1},\sigma_{2})}+\nonumber\\&&  \frac{1}{\cosh^{2}(G(TA_{1},TA_{2},\sigma_{1},\sigma_{2}))}(\frac{dz}{dr_{2}})^{2} \Big) dr_{2}^{2} -\nonumber\\&&  \frac{1}{\sinh(G(TA_{1},TA_{2},\sigma_{1},\sigma_{2}))\cosh(G(TA_{1},TA_{2},\sigma_{1},\sigma_{2}))}(\frac{dz}{dt_{2} }\frac{dz}{dr_{2}})dt_{2} dr_{2} + \nonumber\\&&  D_{IV-A}^{-\frac{1}{2}} H_{IV-A}^{\frac{1}{2}}\Big(\frac{1}{a}e^{-O(TA_{1},TA_{2},\sigma_{1},\sigma_{2})} \cosh(G(TA_{1},TA_{2},\sigma_{1},\sigma_{2}))\Big)^{2}\Big(d\theta^{2} + sin^{2}\theta d\phi^{2}\Big)  +\nonumber\\&&  D_{IV-A}^{-\frac{1}{2}} H_{IV-A}^{-\frac{1}{2}}\sum_{i=1}^{5}dx_{i}^{2}
\label{a14}
\end{eqnarray}

where

\begin{eqnarray}
&& f_{III-A} = 1-\frac{\Big(e^{E(TA_{1,0},TA_{2,0},\sigma_{1,0},\sigma_{2,0})} \cosh(F(TA_{1,0},TA_{2,0},\sigma_{1,0},\sigma_{2,0}))\Big)^{4}}{\Big(e^{F(TA_{1},TA_{2},\sigma_{1},\sigma_{2})} \cosh(F(TA_{1},TA_{2},\sigma_{1},\sigma_{2}))\Big)^{4}} \nonumber\\&& H_{III-A} = 1+\frac{\Big(e^{E(TA_{1,0},TA_{2,0},\sigma_{1,0},\sigma_{2,0})} \cosh(F(TA_{1,0},TA_{2,0},\sigma_{1,0},\sigma_{2,0}))\Big)^{4}\sinh^{2}\alpha_{III-A}}{\Big(e^{E(TA_{1},TA_{2},\sigma_{1},\sigma_{2})} \cosh(F(TA_{1},TA_{2},\sigma_{1},\sigma_{2}))\Big)^{4}}\nonumber\\&& D_{III-A} = \cos^{2}\epsilon_{III-A} + \sin^{2}\epsilon_{III-A} H_{III-A}^{-1}
\label{a15}
\end{eqnarray}

\begin{eqnarray}
&& f_{IV-A} = 1-\frac{\Big(e^{-O(TA_{1,0},TA_{2,0},\sigma_{1,0},\sigma_{2,0})} \cosh(G(TA_{1,0},TA_{2,0},\sigma_{1,0},\sigma_{2,0}))\Big)^{4}}{\Big(e^{-O(TA_{1},TA_{2},\sigma_{1},\sigma_{2})} \cosh(G(TA_{1},TA_{2},\sigma_{1},\sigma_{2}))\Big)^{4}} \nonumber\\&& H_{IV-A} = 1+\frac{\Big(e^{O(TA_{1,0},TA_{2,0},\sigma_{1,0},\sigma_{2,0})} \cosh(G(TA_{1,0},TA_{2,0},\sigma_{1,0},\sigma_{2,0}))\Big)^{4}\sinh^{2}\alpha_{IV-A}}{\Big(e^{O(TA_{1},TA_{2},\sigma_{1},\sigma_{2})} \cosh(G(TA_{1},TA_{2},\sigma_{1},\sigma_{2}))\Big)^{4}}\nonumber\\&& D_{IV-A} = \cos^{2}\epsilon_{IV-A} + \sin^{2}\epsilon_{IV-A} H_{IV-A}^{-1}
\label{a16}
\end{eqnarray}

and

\begin{eqnarray}
&& \cosh^{2} \alpha_{III-A} = \frac{3}{2}\frac{\cos\frac{\delta_{III-A}}{3} + \sqrt{3}\cos\frac{\delta_{III-A}}{3}}{\cos\delta_{III-A}}\nonumber\\&& \cos\epsilon_{III-A} = \frac{1}{\sqrt{1 + \frac{K^{2}}{\Big(a_{1}^{-1}e^{-E(TA_{1,0},TA_{2,0},\sigma_{1,0},\sigma_{2,0})} \cosh(F(TA_{1,0},TA_{2,0},\sigma_{1,0},\sigma_{2,0}))\Big)^{4}}}}
\label{a17}
\end{eqnarray}

\begin{eqnarray}
&& \cosh^{2} \alpha_{IV-A} = \frac{3}{2}\frac{\cos\frac{\delta_{IV-A}}{3} + \sqrt{3}\cos\frac{\delta_{IV-A}}{3}}{\cos\delta_{IV-A}}\nonumber\\&& \cos\epsilon_{IV-A} = \frac{1}{\sqrt{1 + \frac{K^{2}}{\Big(a_{2}^{-1}e^{O(TA_{1,0},TA_{2,0},\sigma_{1,0},\sigma_{2,0})} \cosh(G(TA_{1,0},TA_{2,0},\sigma_{1,0},\sigma_{2,0}))\Big)^{4}}}}
\label{a18}
\end{eqnarray}

The angles $\delta_{I-A}$ and $\delta_{II-A}$ are defined by:

\begin{eqnarray}
&& \cos\delta_{III-A} = \bar{T}_{0,III-A}^{4}\sqrt{1 + \frac{K^{2}}{\Big(a_{1}^{-1}e^{-E(TA_{1},TA_{2},\sigma_{1},\sigma_{2}))} \cosh(F(TA_{1},TA_{2},\sigma_{1},\sigma_{2})))\Big)^{4}}}\nonumber\\&&   \bar{T}_{0,III-A} = \Big(\frac{9\pi^{2}N}{4\sqrt{3}T_{D3}}\Big)^{\frac{1}{2}}T_{0,III-A}
\label{a19}
\end{eqnarray}

\begin{eqnarray}
&& \cos\delta_{IV-A} = \bar{T}_{0,IV-A}^{4}\sqrt{1 + \frac{K^{2}}{\Big(a_{2}^{-1}e^{O(TA_{1},TA_{2},\sigma_{1},\sigma_{2}))} \cosh(G(TA_{1},TA_{2},\sigma_{1},\sigma_{2})))\Big)^{4}}}\nonumber\\&&   \bar{T}_{0,IV-A} = \Big(\frac{9\pi^{2}N}{4\sqrt{3}T_{D3}}\Big)^{\frac{1}{2}}T_{0,IV-A}
\label{a20}
\end{eqnarray}

Here $T_{0}$ is the temperature of the BIon in non-Rindler space-time.  For the above metric, the energy densities and entropies in the two regions are:

\begin{eqnarray}
&& \int d E(TA_{1},TA_{2},\sigma_{1},\sigma_{2}) \frac{d M_{III-A}}{dz}= \nonumber\\&&\frac{4 T_{D3}^{2}}{\pi T_{0,III-A}^{4}} \int_{\sigma_{1,0}}^{\infty}  d E(TA_{1},TA_{2},\sigma_{1},\sigma_{2}) \times \nonumber\\&&  \frac{F_{DBI,III,A}(TA_{1},TA_{2},\sigma_{1},\sigma_{2})\Big(\frac{1}{a}e^{E(TA_{1},TA_{2},\sigma_{1},\sigma_{2})} \cosh(F(TA_{1},TA_{2},\sigma_{1},\sigma_{2}))\Big)^{2}}{\sqrt{F_{DBI,III,A}^{2}(TA_{1},TA_{2},\sigma_{1},\sigma_{2})-F_{DBI,III,A}^{2}(TA_{1},TA_{2},\sigma_{1,0},\sigma_{2})}}\times \nonumber\\&& \frac{[4 \cosh^{2}\alpha_{III-A} + 1]\Big(\sinh^{2}(F(TA_{1},TA_{2},\sigma_{1},\sigma_{2}))+cosh^{2}(F(TA_{1},TA_{2},\sigma_{1},\sigma_{2}))\Big)}{\cosh^{4}\alpha_{III-A}}\nonumber\\&& S_{III-A}= \nonumber\\&&\frac{4 T_{D3}^{2}}{\pi T_{0,III-A}^{5}} \int_{\sigma_{1,0}}^{\infty}  d E(TA_{1},TA_{2},\sigma_{1},\sigma_{2}) \times \nonumber\\&&   \frac{F_{DBI,III,A}(TA_{1},TA_{2},\sigma_{1},\sigma_{2})\Big(\frac{1}{a_{1}}e^{E(TA_{1},TA_{2},\sigma_{1},\sigma_{2})} \cosh(F(TA_{1},TA_{2},\sigma_{1},\sigma_{2}))\Big)^{2}}{\sqrt{F_{DBI,III,A}^{2}(TA_{1},TA_{2},\sigma_{1},\sigma_{2})-F_{DBI,III,A}^{2}(TA_{1},TA_{2},\sigma_{1,0},\sigma_{2})}}\times \nonumber\\&& \frac{4\Big(\sinh^{2}(F(TA_{1},TA_{2},\sigma_{1},\sigma_{2}))+cosh^{2}(F(TA_{1},TA_{2},\sigma_{1},\sigma_{2}))\Big) }{\cosh^{4}\alpha_{III-A}}
\label{a21}
\end{eqnarray}

with the definition of $F_{DBI,III,A}$ given below:

\begin{eqnarray}
&& F_{DBI,III,A} = \Big(a_{1}^{-1}e^{E(TA_{1},TA_{2},\sigma_{1},\sigma_{2})} \cosh(F(TA_{1},TA_{2},\sigma_{1},\sigma_{2}))\Big)^{2}\frac{4\cosh^{2}\alpha_{III-A} - 3}{\cosh^{4}\alpha_{III-A}}
\label{a22}
\end{eqnarray}

and

\begin{eqnarray}
&&  \int d O(TA_{1},TA_{2},\sigma_{1},\sigma_{2})  \frac{d M_{IV-A}}{dz}=   \nonumber\\&&\frac{4 T_{D3}^{2}}{\pi T_{0,IV-A}^{4}} \int_{\sigma_{2,0}}^{\infty}  d O(TA_{1},TA_{2},\sigma_{1},\sigma_{2})\times \nonumber\\&&   \frac{F_{DBI,IV,A}(TA_{1},TA_{2},\sigma_{1},\sigma_{2,0})\Big(\frac{1}{a_{2}}e^{-O(TA_{1},TA_{2},\sigma_{1},\sigma_{2})} \cosh(G(TA_{1},TA_{2},\sigma_{1},\sigma_{2}))\Big)^{2}}{\sqrt{F_{DBI,II,A}^{2}(TA_{1},TA_{2},\sigma_{1},\sigma_{2})-F_{DBI,IV,A}^{2}(TA_{1},TA_{2},\sigma_{1},\sigma_{2,0}))}}\times \nonumber\\&& \frac{[4 \cosh^{2}\alpha_{IV-A} + 1][\Big(\sinh^{2}(G(TA_{1},TA_{2},\sigma_{1},\sigma_{2}))+cosh^{2}(G(TA_{1},TA_{2},\sigma_{1},\sigma_{2}))\Big)]}{\cosh^{4}\alpha_{IV-A}}\nonumber\\&& S_{IV-A}= \nonumber\\&&\frac{4 T_{D3}^{2}}{\pi T_{0,IV-A}^{5}} \int_{\sigma_{2,0}}^{\infty}  d O(TA_{1},TA_{2},\sigma_{1},\sigma_{2})\times \nonumber\\&& \frac{F_{DBI,IV,A}(TA_{1},TA_{2},\sigma_{1},\sigma_{2})\Big(\frac{1}{a_{2}}e^{-O(TA_{1},TA_{2},\sigma_{1},\sigma_{2})} \cosh(G(TA_{1},TA_{2},\sigma_{1},\sigma_{2}))\Big)^{2}}{\sqrt{F_{DBI,IV,A}^{2}(TA_{1},TA_{2},\sigma_{1},\sigma_{2})-F_{DBI,IV,A}^{2}(TA_{1},TA_{2},\sigma_{1},\sigma_{2,0})}}\times \nonumber\\&& \frac{4\Big(\sinh^{2}(G(TA_{1},TA_{2},\sigma_{1},\sigma_{2}))+cosh^{2}(G(TA_{1},TA_{2},\sigma_{1},\sigma_{2}))\Big) }{\cosh^{4}\alpha_{IV-A}}\label{a23}
\end{eqnarray}

with the  definition of $F_{DBI,IV,A}$ given below:

\begin{eqnarray}
&& F_{DBI,IV,A} = \Big(a_{2}^{-1}e^{-O(TA_{1},TA_{2},\sigma_{1},\sigma_{2})} \cosh(G(TA_{1},TA_{2},\sigma_{1},\sigma_{2}))\Big)^{2}\frac{4\cosh^{2}\alpha_{IV-A} - 3}{\cosh^{4}\alpha_{IV-A}}
\label{a24}
\end{eqnarray}

The above results show that the interaction between two tachyonic wormholes changes the thermal properties of the system in the two regions of space-time. The thermal parameters like entropy and energy density depend on the tachyonic fields, the separation between the branes, the coordinates of the system and the size of the throats of the  wormholes.  Also,  the entropy and energy densities of the wormholes in one region acts in opposite order  to the ones in the other region. In particular, when the  entropy in one region increases, the entropy in the other region decreases. 

\section{ Padmanabhan mechanism in the background of two interacting tachyonic BIons}\label{o2}

In this section, we build the Padmanabhan mechanism in a BIon with two tachyonic wormholes. In this system, four universes emerge which live in four regions of the BIon (See figure 2). Two universes have been considered in \cite{q8,q12}. Now, we consider the behaviour of the two universes in the two new regions, which are emerged from the interaction of the wormholes. Previously, it was shown that the number of degrees of freedom on the surface and within the bulk have the  relations below  with the entropy and mass densities \cite{q6,q7,q8}:

\begin{eqnarray}
&& N_{sur,III-A}-N_{bulk,III-A}=\int d E(TA_{1},TA_{2},\sigma_{1},\sigma_{2}) \frac{d M_{III-A}}{dz}=\nonumber\\&&  N_{sur,I-A}+N_{bulk,I-A}=\frac{4\pi}{L_{P}^{2}}S_{III-A}
\label{a25}
\end{eqnarray}

\begin{eqnarray}
&& N_{sur,IV-A}-N_{bulk,IV-A}=\int d O(TA_{1},TA_{2},\sigma_{1},\sigma_{2}) \frac{d M_{IV-A}}{dz}= \nonumber\\&&  N_{sur,IV-A}+N_{bulk,IV-A}=\frac{4\pi}{L_{P}^{2}}S_{IV-A}
\label{a26}
\end{eqnarray}

On the other hand, previously, it was shown that the Hubble parameter of the universe has the following relation with the number of degrees of freedom on the surface of the BIon \cite{q8}:

\begin{eqnarray}
&&  N_{sur} = \frac{4 \pi r_{A}^{2}}{L_{P}^{2}}\nonumber\\&& r_{A}=\sqrt{H^{2} + \frac{k^{2}}{\bar{a}^{2}}}
\label{a27}
\end{eqnarray}
where $H = \frac{\bar{a}}{\dot{\bar{a}}}$ is the Hubble parameter, $\bar{a}$ is the scale factor and $r_{A}$ is the apparent horizon radius for the FLRW universe. Thus, using equations (\ref{a25}, \ref{a26},\ref{a27}) and assuming $k=0$, we obtain:

\begin{eqnarray}
&& N_{sur,III-A}=\frac{1}{2}[\int d E(TA_{1},TA_{2},\sigma_{1},\sigma_{2})\frac{d M_{III-A}}{dz} + \frac{4\pi}{L_{P}^{2}}S_{III-A}
]= \frac{4 \pi H_{III-A}^{2}}{L_{P}^{2}}\label{a28}
\end{eqnarray}

\begin{eqnarray}
&& N_{sur,IV-A}=\frac{1}{2}[\int d O(TA_{1},TA_{2},\sigma_{1},\sigma_{2}) \frac{d M_{IV-A}}{dz} + \frac{4\pi}{L_{P}^{2}}S_{IV-A}
]= \frac{4 \pi H_{IV-A}^{2}}{L_{P}^{2}}\label{a29}
\end{eqnarray}

Thus, the Hubble parameters in the two regions can be obtained as:

\begin{eqnarray}
&& H_{III-A}=\sqrt{\frac{L_{P}^{2}}{4 \pi}}\times \nonumber\\&&[\frac{1}{2}[\frac{4 T_{D3}^{2}}{\pi T_{0,III-A}^{4}} \int_{\sigma_{1,0}}^{\infty}  d E(TA_{1},TA_{2},\sigma_{1},\sigma_{2}) \times \nonumber\\&&  \frac{F_{DBI,III,A}(TA_{1},TA_{2},\sigma_{1},\sigma_{2})\Big(\frac{1}{a}e^{E(TA_{1},TA_{2},\sigma_{1},\sigma_{2})} \cosh(F(TA_{1},TA_{2},\sigma_{1},\sigma_{2}))\Big)^{2}}{\sqrt{F_{DBI,III,A}^{2}(TA_{1},TA_{2},\sigma_{1},\sigma_{2})-F_{DBI,III,A}^{2}(TA_{1},TA_{2},\sigma_{1,0},\sigma_{2})}}\times \nonumber\\&& \frac{[4 \cosh^{2}\alpha_{III-A} + 5]\Big(\sinh^{2}(F(TA_{1},TA_{2},\sigma_{1},\sigma_{2}))+cosh^{2}(F(TA_{1},TA_{2},\sigma_{1},\sigma_{2}))\Big)}{\cosh^{4}\alpha_{III-A}} 
]]^{\frac{1}{2}} \label{a30}
\end{eqnarray}

\begin{eqnarray}
&& H_{IV-A}=\sqrt{\frac{L_{P}^{2}}{4 \pi}}\times \nonumber\\&&[\frac{1}{2}[\frac{4 T_{D3}^{2}}{\pi T_{0,IV-A}^{4}} \int_{\sigma_{2,0}}^{\infty}  d O(TA_{1},TA_{2},\sigma_{1},\sigma_{2})\times \nonumber\\&&   \frac{F_{DBI,IV,A}(TA_{1},TA_{2},\sigma_{1},\sigma_{2,0})\Big(\frac{1}{a_{2}}e^{-O(TA_{1},TA_{2},\sigma_{1},\sigma_{2})} \cosh(G(TA_{1},TA_{2},\sigma_{1},\sigma_{2}))\Big)^{2}}{\sqrt{F_{DBI,II,A}^{2}(TA_{1},TA_{2},\sigma_{1},\sigma_{2})-F_{DBI,IV,A}^{2}(TA_{1},TA_{2},\sigma_{1},\sigma_{2,0}))}}\times \nonumber\\&& \frac{[4 \cosh^{2}\alpha_{IV-A} + 5][\Big(\sinh^{2}(G(TA_{1},TA_{2},\sigma_{1},\sigma_{2}))+cosh^{2}(G(TA_{1},TA_{2},\sigma_{1},\sigma_{2}))\Big)]}{\cosh^{4}\alpha_{IV-A}}
]]^{\frac{1}{2}}= \label{a31}
\end{eqnarray}

Substituting the results from equation ( \ref{a12}) in equations (\ref{a30} and \ref{a31}), we obtain:

\begin{eqnarray}
&& H_{III-A}\sim \sqrt{\frac{L_{P}^{2}}{4 \pi}}[\frac{1}{2}[\frac{4 T_{D3}^{2}}{\pi T_{0,III-A}^{4}} \times \nonumber\\&&  [\frac{V(TA_{1})\sqrt{1 + \frac{l'(\sigma_{1})^{2}}{4}+ TA_{1}^{2}l_{1}^{2}}\sigma_{1}^{2}\sqrt{1 + \frac{\beta}{\sigma_{1}^{4}}}}{m_{TA_{1}}}]^{-2} \times \nonumber\\&& exp[4[ e^{\frac{\frac{V(TA_{2})\sqrt{1 + \frac{l'(\sigma_{2})^{2}}{4}+ TA_{2}^{2}l_{2}^{2}}\sigma_{2}^{2}\sqrt{1 + \frac{\beta}{\sigma_{2}^{4}}}}{m_{TA_{2}}}}{\frac{V(TA_{1})\sqrt{1 + \frac{l'(\sigma_{1})^{2}}{4}+ TA_{1}^{2}l_{1}^{2}}\sigma_{1}^{2}\sqrt{1 + \frac{\beta}{\sigma_{1}^{4}}}}{m_{TA_{1}}}}e^{\frac{V(TA_{1})\sqrt{1 + \frac{l'(\sigma_{1})^{2}}{4}+ TA_{1}^{2}l_{1}^{2}}\sigma_{1}^{2}\sqrt{1 + \frac{\beta}{\sigma_{1}^{4}}}}{m_{TA_{1}}}\sigma_{1}} \cosh(\frac{V(TA_{1})\sqrt{1 + \frac{l'(\sigma_{1})^{2}}{4}+ TA_{1}^{2}l_{1}^{2}}\sigma_{1}^{2}\sqrt{1 + \frac{\beta}{\sigma_{1}^{4}}}}{m_{TA_{1}}}\tau_{1})} \times \nonumber\\&&\cosh(\frac{\frac{V(TA_{2})\sqrt{1 + \frac{l'(\sigma_{2})^{2}}{4}+ TA_{2}^{2}l_{2}^{2}}\sigma_{2}^{2}\sqrt{1 + \frac{\beta}{\sigma_{2}^{4}}}}{m_{TA_{2}}}}{\frac{V(TA_{1})\sqrt{1 + \frac{l'(\sigma_{1})^{2}}{4}+ TA_{1}^{2}l_{1}^{2}}\sigma_{1}^{2}\sqrt{1 + \frac{\beta}{\sigma_{1}^{4}}}}{m_{TA_{1}}}}e^{\frac{V(TA_{1})\sqrt{1 + \frac{l'(\sigma_{1})^{2}}{4}+ TA_{1}^{2}l_{1}^{2}}\sigma_{1}^{2}\sqrt{1 + \frac{\beta}{\sigma_{1}^{4}}}}{m_{TA_{1}}}\sigma_{1}}\times \nonumber\\&& \sinh(\frac{V(TA_{1})\sqrt{1 + \frac{l'(\sigma_{1})^{2}}{4}+ TA_{1}^{2}l_{1}^{2}}\sigma_{1}^{2}\sqrt{1 + \frac{\beta}{\sigma_{1}^{4}}}}{m_{TA_{1}}}\tau_{1}))]]\times \nonumber\\&&\Big( \cosh(e^{\frac{\frac{V(TA_{2})\sqrt{1 + \frac{l'(\sigma_{2})^{2}}{4}+ TA_{2}^{2}l_{2}^{2}}\sigma_{2}^{2}\sqrt{1 + \frac{\beta}{\sigma_{2}^{4}}}}{m_{TA_{2}}}}{\frac{V(TA_{1})\sqrt{1 + \frac{l'(\sigma_{1})^{2}}{4}+ TA_{1}^{2}l_{1}^{2}}\sigma_{1}^{2}\sqrt{1 + \frac{\beta}{\sigma_{1}^{4}}}}{m_{TA_{1}}}}e^{\frac{V(TA_{1})\sqrt{1 + \frac{l'(\sigma_{1})^{2}}{4}+ TA_{1}^{2}l_{1}^{2}}\sigma_{1}^{2}\sqrt{1 + \frac{\beta}{\sigma_{1}^{4}}}}{m_{TA_{1}}}\sigma_{1}} \cosh(\frac{V(TA_{1})\sqrt{1 + \frac{l'(\sigma_{1})^{2}}{4}+ TA_{1}^{2}l_{1}^{2}}\sigma_{1}^{2}\sqrt{1 + \frac{\beta}{\sigma_{1}^{4}}}}{m_{TA_{1}}}\tau_{1})} \times \nonumber\\&& \sinh(\frac{\frac{V(TA_{2})\sqrt{1 + \frac{l'(\sigma_{2})^{2}}{4}+ TA_{2}^{2}l_{2}^{2}}\sigma_{2}^{2}\sqrt{1 + \frac{\beta}{\sigma_{2}^{4}}}}{m_{TA_{2}}}}{\frac{V(TA_{1})\sqrt{1 + \frac{l'(\sigma_{1})^{2}}{4}+ TA_{1}^{2}l_{1}^{2}}\sigma_{1}^{2}\sqrt{1 + \frac{\beta}{\sigma_{1}^{4}}}}{m_{TA_{1}}}}e^{\frac{V(TA_{1})\sqrt{1 + \frac{l'(\sigma_{1})^{2}}{4}+ TA_{1}^{2}l_{1}^{2}}\sigma_{1}^{2}\sqrt{1 + \frac{\beta}{\sigma_{1}^{4}}}}{m_{TA_{1}}}\sigma_{1}} \times \nonumber\\&&\sinh(\frac{V(TA_{1})\sqrt{1 + \frac{l'(\sigma_{1})^{2}}{4}+ TA_{1}^{2}l_{1}^{2}}\sigma_{1}^{2}\sqrt{1 + \frac{\beta}{\sigma_{1}^{4}}}}{m_{TA_{1}}}\tau_{1})))\Big)^{2}\times\nonumber\\&& \Big(\sinh^{2}(e^{\frac{\frac{V(TA_{2})\sqrt{1 + \frac{l'(\sigma_{2})^{2}}{4}+ TA_{2}^{2}l_{2}^{2}}\sigma_{2}^{2}\sqrt{1 + \frac{\beta}{\sigma_{2}^{4}}}}{m_{TA_{2}}}}{\frac{V(TA_{1})\sqrt{1 + \frac{l'(\sigma_{1})^{2}}{4}+ TA_{1}^{2}l_{1}^{2}}\sigma_{1}^{2}\sqrt{1 + \frac{\beta}{\sigma_{1}^{4}}}}{m_{TA_{1}}}}e^{\frac{V(TA_{1})\sqrt{1 + \frac{l'(\sigma_{1})^{2}}{4}+ TA_{1}^{2}l_{1}^{2}}\sigma_{1}^{2}\sqrt{1 + \frac{\beta}{\sigma_{1}^{4}}}}{m_{TA_{1}}}\sigma_{1}} \cosh(\frac{V(TA_{1})\sqrt{1 + \frac{l'(\sigma_{1})^{2}}{4}+ TA_{1}^{2}l_{1}^{2}}\sigma_{1}^{2}\sqrt{1 + \frac{\beta}{\sigma_{1}^{4}}}}{m_{TA_{1}}}\tau_{1})} \times \nonumber\\&& \sinh(\frac{\frac{V(TA_{2})\sqrt{1 + \frac{l'(\sigma_{2})^{2}}{4}+ TA_{2}^{2}l_{2}^{2}}\sigma_{2}^{2}\sqrt{1 + \frac{\beta}{\sigma_{2}^{4}}}}{m_{TA_{2}}}}{\frac{V(TA_{1})\sqrt{1 + \frac{l'(\sigma_{1})^{2}}{4}+ TA_{1}^{2}l_{1}^{2}}\sigma_{1}^{2}\sqrt{1 + \frac{\beta}{\sigma_{1}^{4}}}}{m_{TA_{1}}}}e^{\frac{V(TA_{1})\sqrt{1 + \frac{l'(\sigma_{1})^{2}}{4}+ TA_{1}^{2}l_{1}^{2}}\sigma_{1}^{2}\sqrt{1 + \frac{\beta}{\sigma_{1}^{4}}}}{m_{TA_{1}}}\sigma_{1}} \times \nonumber\\&&\sinh(\frac{V(TA_{1})\sqrt{1 + \frac{l'(\sigma_{1})^{2}}{4}+ TA_{1}^{2}l_{1}^{2}}\sigma_{1}^{2}\sqrt{1 + \frac{\beta}{\sigma_{1}^{4}}}}{m_{TA_{1}}}\tau_{1})))+ \nonumber\\&& cosh^{2}(e^{\frac{\frac{V(TA_{2})\sqrt{1 + \frac{l'(\sigma_{2})^{2}}{4}+ TA_{2}^{2}l_{2}^{2}}\sigma_{2}^{2}\sqrt{1 + \frac{\beta}{\sigma_{2}^{4}}}}{m_{TA_{2}}}}{\frac{V(TA_{1})\sqrt{1 + \frac{l'(\sigma_{1})^{2}}{4}+ TA_{1}^{2}l_{1}^{2}}\sigma_{1}^{2}\sqrt{1 + \frac{\beta}{\sigma_{1}^{4}}}}{m_{TA_{1}}}}e^{\frac{V(TA_{1})\sqrt{1 + \frac{l'(\sigma_{1})^{2}}{4}+ TA_{1}^{2}l_{1}^{2}}\sigma_{1}^{2}\sqrt{1 + \frac{\beta}{\sigma_{1}^{4}}}}{m_{TA_{1}}}\sigma_{1}} \cosh(\frac{V(TA_{1})\sqrt{1 + \frac{l'(\sigma_{1})^{2}}{4}+ TA_{1}^{2}l_{1}^{2}}\sigma_{1}^{2}\sqrt{1 + \frac{\beta}{\sigma_{1}^{4}}}}{m_{TA_{1}}}\tau_{1})} \times \nonumber\\&& \sinh(\frac{\frac{V(TA_{2})\sqrt{1 + \frac{l'(\sigma_{2})^{2}}{4}+ TA_{2}^{2}l_{2}^{2}}\sigma_{2}^{2}\sqrt{1 + \frac{\beta}{\sigma_{2}^{4}}}}{m_{TA_{2}}}}{\frac{V(TA_{1})\sqrt{1 + \frac{l'(\sigma_{1})^{2}}{4}+ TA_{1}^{2}l_{1}^{2}}\sigma_{1}^{2}\sqrt{1 + \frac{\beta}{\sigma_{1}^{4}}}}{m_{TA_{1}}}}e^{\frac{V(TA_{1})\sqrt{1 + \frac{l'(\sigma_{1})^{2}}{4}+ TA_{1}^{2}l_{1}^{2}}\sigma_{1}^{2}\sqrt{1 + \frac{\beta}{\sigma_{1}^{4}}}}{m_{TA_{1}}}\sigma_{1}} \times \nonumber\\&&\sinh(\frac{V(TA_{1})\sqrt{1 + \frac{l'(\sigma_{1})^{2}}{4}+ TA_{1}^{2}l_{1}^{2}}\sigma_{1}^{2}\sqrt{1 + \frac{\beta}{\sigma_{1}^{4}}}}{m_{TA_{1}}}\tau_{1})))\Big)\times\nonumber\\&& [4 \cosh^{2}\alpha_{III-A} + 5]\times \nonumber\\&&\cosh^{-4}\alpha_{III-A} \label{a32}
\end{eqnarray}

and

\begin{eqnarray}
&& H_{IV-A}\sim \sqrt{\frac{L_{P}^{2}}{4 \pi}}[\frac{1}{2}[\frac{4 T_{D3}^{2}}{\pi T_{0,IV-A}^{4}} \times \nonumber\\&&  [\frac{V(TA_{2})\sqrt{1 + \frac{l'(\sigma_{2})^{2}}{4}+ TA_{2}^{2}l_{2}^{2}}\sigma_{2}^{2}\sqrt{1 + \frac{\beta}{\sigma_{2}^{4}}}}{m_{TA_{2}}}]^{-2} \times \nonumber\\&& exp[4[E(TA_{1},TA_{2},\sigma_{1},\sigma_{2})]]\times \nonumber\\&&\Big( \cosh(- e^{-\frac{\frac{V(TA_{2})\sqrt{1 + \frac{l'(\sigma_{2})^{2}}{4}+ TA_{2}^{2}l_{2}^{2}}\sigma_{2}^{2}\sqrt{1 + \frac{\beta}{\sigma_{2}^{4}}}}{m_{TA_{2}}}}{\frac{V(TA_{1})\sqrt{1 + \frac{l'(\sigma_{1})^{2}}{4}+ TA_{1}^{2}l_{1}^{2}}\sigma_{1}^{2}\sqrt{1 + \frac{\beta}{\sigma_{1}^{4}}}}{m_{TA_{1}}}}e^{-a_{1}\sigma_{1}} \cosh(a_{1}\tau_{1})} \times \nonumber\\&&\sinh(-\frac{\frac{V(TA_{2})\sqrt{1 + \frac{l'(\sigma_{2})^{2}}{4}+ TA_{2}^{2}l_{2}^{2}}\sigma_{2}^{2}\sqrt{1 + \frac{\beta}{\sigma_{2}^{4}}}}{m_{TA_{2}}}}{\frac{V(TA_{1})\sqrt{1 + \frac{l'(\sigma_{1})^{2}}{4}+ TA_{1}^{2}l_{1}^{2}}\sigma_{1}^{2}\sqrt{1 + \frac{\beta}{\sigma_{1}^{4}}}}{m_{TA_{1}}}}e^{-\frac{V(TA_{1})\sqrt{1 + \frac{l'(\sigma_{1})^{2}}{4}+ TA_{1}^{2}l_{1}^{2}}\sigma_{1}^{2}\sqrt{1 + \frac{\beta}{\sigma_{1}^{4}}}}{m_{TA_{1}}}\sigma_{1}}\times \nonumber\\&& \sinh(\frac{V(TA_{1})\sqrt{1 + \frac{l'(\sigma_{1})^{2}}{4}+ TA_{1}^{2}l_{1}^{2}}\sigma_{1}^{2}\sqrt{1 + \frac{\beta}{\sigma_{1}^{4}}}}{m_{TA_{1}}}\tau_{1})))\Big)^{2}\times\nonumber\\&&\Big(\sinh^{2}(- e^{-\frac{\frac{V(TA_{2})\sqrt{1 + \frac{l'(\sigma_{2})^{2}}{4}+ TA_{2}^{2}l_{2}^{2}}\sigma_{2}^{2}\sqrt{1 + \frac{\beta}{\sigma_{2}^{4}}}}{m_{TA_{2}}}}{\frac{V(TA_{1})\sqrt{1 + \frac{l'(\sigma_{1})^{2}}{4}+ TA_{1}^{2}l_{1}^{2}}\sigma_{1}^{2}\sqrt{1 + \frac{\beta}{\sigma_{1}^{4}}}}{m_{TA_{1}}}}e^{-a_{1}\sigma_{1}} \cosh(a_{1}\tau_{1})} \times \nonumber\\&&\sinh(-\frac{\frac{V(TA_{2})\sqrt{1 + \frac{l'(\sigma_{2})^{2}}{4}+ TA_{2}^{2}l_{2}^{2}}\sigma_{2}^{2}\sqrt{1 + \frac{\beta}{\sigma_{2}^{4}}}}{m_{TA_{2}}}}{\frac{V(TA_{1})\sqrt{1 + \frac{l'(\sigma_{1})^{2}}{4}+ TA_{1}^{2}l_{1}^{2}}\sigma_{1}^{2}\sqrt{1 + \frac{\beta}{\sigma_{1}^{4}}}}{m_{TA_{1}}}}e^{-\frac{V(TA_{1})\sqrt{1 + \frac{l'(\sigma_{1})^{2}}{4}+ TA_{1}^{2}l_{1}^{2}}\sigma_{1}^{2}\sqrt{1 + \frac{\beta}{\sigma_{1}^{4}}}}{m_{TA_{1}}}\sigma_{1}}\times \nonumber\\&& \sinh(\frac{V(TA_{1})\sqrt{1 + \frac{l'(\sigma_{1})^{2}}{4}+ TA_{1}^{2}l_{1}^{2}}\sigma_{1}^{2}\sqrt{1 + \frac{\beta}{\sigma_{1}^{4}}}}{m_{TA_{1}}}\tau_{1})))+ \nonumber\\&& cosh^{2}(- e^{-\frac{\frac{V(TA_{2})\sqrt{1 + \frac{l'(\sigma_{2})^{2}}{4}+ TA_{2}^{2}l_{2}^{2}}\sigma_{2}^{2}\sqrt{1 + \frac{\beta}{\sigma_{2}^{4}}}}{m_{TA_{2}}}}{\frac{V(TA_{1})\sqrt{1 + \frac{l'(\sigma_{1})^{2}}{4}+ TA_{1}^{2}l_{1}^{2}}\sigma_{1}^{2}\sqrt{1 + \frac{\beta}{\sigma_{1}^{4}}}}{m_{TA_{1}}}}e^{-a_{1}\sigma_{1}} \cosh(a_{1}\tau_{1})} \times \nonumber\\&&\sinh(-\frac{\frac{V(TA_{2})\sqrt{1 + \frac{l'(\sigma_{2})^{2}}{4}+ TA_{2}^{2}l_{2}^{2}}\sigma_{2}^{2}\sqrt{1 + \frac{\beta}{\sigma_{2}^{4}}}}{m_{TA_{2}}}}{\frac{V(TA_{1})\sqrt{1 + \frac{l'(\sigma_{1})^{2}}{4}+ TA_{1}^{2}l_{1}^{2}}\sigma_{1}^{2}\sqrt{1 + \frac{\beta}{\sigma_{1}^{4}}}}{m_{TA_{1}}}}e^{-\frac{V(TA_{1})\sqrt{1 + \frac{l'(\sigma_{1})^{2}}{4}+ TA_{1}^{2}l_{1}^{2}}\sigma_{1}^{2}\sqrt{1 + \frac{\beta}{\sigma_{1}^{4}}}}{m_{TA_{1}}}\sigma_{1}}\times \nonumber\\&& \sinh(\frac{V(TA_{1})\sqrt{1 + \frac{l'(\sigma_{1})^{2}}{4}+ TA_{1}^{2}l_{1}^{2}}\sigma_{1}^{2}\sqrt{1 + \frac{\beta}{\sigma_{1}^{4}}}}{m_{TA_{1}}}\tau_{1})))\Big)\times\nonumber\\&& [4 \cosh^{2}\alpha_{IV-A} + 5]\times \nonumber\\&&\cosh^{-4}\alpha_{IV-A} \label{a33}
\end{eqnarray}

The above Hubble parameters depend on the tachyonic fields, the brane coordinates, the size of the throat of the wormholes and the separation  between the branes. They show that by the expandion of the universe in one region, the universe in the other region contracts. Also, the universes oscillate between contracting and expanding phases. Using equations (\ref{a32},\ref{a33}) and the mechanism in \cite{q6,q7,q8}, we can obtain the energy density of the system:

\begin{eqnarray}
&& \rho_{III-A}=\frac{3}{8\pi L_{P}^{2}} H_{III-A}^{2} = \nonumber\\&& \frac{3}{8\pi L_{P}^{2}}[\sqrt{\frac{L_{P}^{2}}{4 \pi}}[\frac{1}{2}[\frac{4 T_{D3}^{2}}{\pi T_{0,III-A}^{4}} \times \nonumber\\&&  [\frac{V(TA_{1})\sqrt{1 + \frac{l'(\sigma_{1})^{2}}{4}+ TA_{1}^{2}l_{1}^{2}}\sigma_{1}^{2}\sqrt{1 + \frac{\beta}{\sigma_{1}^{4}}}}{m_{TA_{1}}}]^{-2} \times \nonumber\\&& exp[4[ e^{\frac{\frac{V(TA_{2})\sqrt{1 + \frac{l'(\sigma_{2})^{2}}{4}+ TA_{2}^{2}l_{2}^{2}}\sigma_{2}^{2}\sqrt{1 + \frac{\beta}{\sigma_{2}^{4}}}}{m_{TA_{2}}}}{\frac{V(TA_{1})\sqrt{1 + \frac{l'(\sigma_{1})^{2}}{4}+ TA_{1}^{2}l_{1}^{2}}\sigma_{1}^{2}\sqrt{1 + \frac{\beta}{\sigma_{1}^{4}}}}{m_{TA_{1}}}}e^{\frac{V(TA_{1})\sqrt{1 + \frac{l'(\sigma_{1})^{2}}{4}+ TA_{1}^{2}l_{1}^{2}}\sigma_{1}^{2}\sqrt{1 + \frac{\beta}{\sigma_{1}^{4}}}}{m_{TA_{1}}}\sigma_{1}} \cosh(\frac{V(TA_{1})\sqrt{1 + \frac{l'(\sigma_{1})^{2}}{4}+ TA_{1}^{2}l_{1}^{2}}\sigma_{1}^{2}\sqrt{1 + \frac{\beta}{\sigma_{1}^{4}}}}{m_{TA_{1}}}\tau_{1})} \times \nonumber\\&&\cosh(\frac{\frac{V(TA_{2})\sqrt{1 + \frac{l'(\sigma_{2})^{2}}{4}+ TA_{2}^{2}l_{2}^{2}}\sigma_{2}^{2}\sqrt{1 + \frac{\beta}{\sigma_{2}^{4}}}}{m_{TA_{2}}}}{\frac{V(TA_{1})\sqrt{1 + \frac{l'(\sigma_{1})^{2}}{4}+ TA_{1}^{2}l_{1}^{2}}\sigma_{1}^{2}\sqrt{1 + \frac{\beta}{\sigma_{1}^{4}}}}{m_{TA_{1}}}}e^{\frac{V(TA_{1})\sqrt{1 + \frac{l'(\sigma_{1})^{2}}{4}+ TA_{1}^{2}l_{1}^{2}}\sigma_{1}^{2}\sqrt{1 + \frac{\beta}{\sigma_{1}^{4}}}}{m_{TA_{1}}}\sigma_{1}}\times \nonumber\\&& \sinh(\frac{V(TA_{1})\sqrt{1 + \frac{l'(\sigma_{1})^{2}}{4}+ TA_{1}^{2}l_{1}^{2}}\sigma_{1}^{2}\sqrt{1 + \frac{\beta}{\sigma_{1}^{4}}}}{m_{TA_{1}}}\tau_{1}))]]\times \nonumber\\&&\Big( \cosh(e^{\frac{\frac{V(TA_{2})\sqrt{1 + \frac{l'(\sigma_{2})^{2}}{4}+ TA_{2}^{2}l_{2}^{2}}\sigma_{2}^{2}\sqrt{1 + \frac{\beta}{\sigma_{2}^{4}}}}{m_{TA_{2}}}}{\frac{V(TA_{1})\sqrt{1 + \frac{l'(\sigma_{1})^{2}}{4}+ TA_{1}^{2}l_{1}^{2}}\sigma_{1}^{2}\sqrt{1 + \frac{\beta}{\sigma_{1}^{4}}}}{m_{TA_{1}}}}e^{\frac{V(TA_{1})\sqrt{1 + \frac{l'(\sigma_{1})^{2}}{4}+ TA_{1}^{2}l_{1}^{2}}\sigma_{1}^{2}\sqrt{1 + \frac{\beta}{\sigma_{1}^{4}}}}{m_{TA_{1}}}\sigma_{1}} \cosh(\frac{V(TA_{1})\sqrt{1 + \frac{l'(\sigma_{1})^{2}}{4}+ TA_{1}^{2}l_{1}^{2}}\sigma_{1}^{2}\sqrt{1 + \frac{\beta}{\sigma_{1}^{4}}}}{m_{TA_{1}}}\tau_{1})} \times \nonumber\\&& \sinh(\frac{\frac{V(TA_{2})\sqrt{1 + \frac{l'(\sigma_{2})^{2}}{4}+ TA_{2}^{2}l_{2}^{2}}\sigma_{2}^{2}\sqrt{1 + \frac{\beta}{\sigma_{2}^{4}}}}{m_{TA_{2}}}}{\frac{V(TA_{1})\sqrt{1 + \frac{l'(\sigma_{1})^{2}}{4}+ TA_{1}^{2}l_{1}^{2}}\sigma_{1}^{2}\sqrt{1 + \frac{\beta}{\sigma_{1}^{4}}}}{m_{TA_{1}}}}e^{\frac{V(TA_{1})\sqrt{1 + \frac{l'(\sigma_{1})^{2}}{4}+ TA_{1}^{2}l_{1}^{2}}\sigma_{1}^{2}\sqrt{1 + \frac{\beta}{\sigma_{1}^{4}}}}{m_{TA_{1}}}\sigma_{1}} \times \nonumber\\&&\sinh(\frac{V(TA_{1})\sqrt{1 + \frac{l'(\sigma_{1})^{2}}{4}+ TA_{1}^{2}l_{1}^{2}}\sigma_{1}^{2}\sqrt{1 + \frac{\beta}{\sigma_{1}^{4}}}}{m_{TA_{1}}}\tau_{1})))\Big)^{2}\times\nonumber\\&& \Big(\sinh^{2}(e^{\frac{\frac{V(TA_{2})\sqrt{1 + \frac{l'(\sigma_{2})^{2}}{4}+ TA_{2}^{2}l_{2}^{2}}\sigma_{2}^{2}\sqrt{1 + \frac{\beta}{\sigma_{2}^{4}}}}{m_{TA_{2}}}}{\frac{V(TA_{1})\sqrt{1 + \frac{l'(\sigma_{1})^{2}}{4}+ TA_{1}^{2}l_{1}^{2}}\sigma_{1}^{2}\sqrt{1 + \frac{\beta}{\sigma_{1}^{4}}}}{m_{TA_{1}}}}e^{\frac{V(TA_{1})\sqrt{1 + \frac{l'(\sigma_{1})^{2}}{4}+ TA_{1}^{2}l_{1}^{2}}\sigma_{1}^{2}\sqrt{1 + \frac{\beta}{\sigma_{1}^{4}}}}{m_{TA_{1}}}\sigma_{1}} \cosh(\frac{V(TA_{1})\sqrt{1 + \frac{l'(\sigma_{1})^{2}}{4}+ TA_{1}^{2}l_{1}^{2}}\sigma_{1}^{2}\sqrt{1 + \frac{\beta}{\sigma_{1}^{4}}}}{m_{TA_{1}}}\tau_{1})} \times \nonumber\\&& \sinh(\frac{\frac{V(TA_{2})\sqrt{1 + \frac{l'(\sigma_{2})^{2}}{4}+ TA_{2}^{2}l_{2}^{2}}\sigma_{2}^{2}\sqrt{1 + \frac{\beta}{\sigma_{2}^{4}}}}{m_{TA_{2}}}}{\frac{V(TA_{1})\sqrt{1 + \frac{l'(\sigma_{1})^{2}}{4}+ TA_{1}^{2}l_{1}^{2}}\sigma_{1}^{2}\sqrt{1 + \frac{\beta}{\sigma_{1}^{4}}}}{m_{TA_{1}}}}e^{\frac{V(TA_{1})\sqrt{1 + \frac{l'(\sigma_{1})^{2}}{4}+ TA_{1}^{2}l_{1}^{2}}\sigma_{1}^{2}\sqrt{1 + \frac{\beta}{\sigma_{1}^{4}}}}{m_{TA_{1}}}\sigma_{1}} \times \nonumber\\&&\sinh(\frac{V(TA_{1})\sqrt{1 + \frac{l'(\sigma_{1})^{2}}{4}+ TA_{1}^{2}l_{1}^{2}}\sigma_{1}^{2}\sqrt{1 + \frac{\beta}{\sigma_{1}^{4}}}}{m_{TA_{1}}}\tau_{1})))+ \nonumber\\&& cosh^{2}(e^{\frac{\frac{V(TA_{2})\sqrt{1 + \frac{l'(\sigma_{2})^{2}}{4}+ TA_{2}^{2}l_{2}^{2}}\sigma_{2}^{2}\sqrt{1 + \frac{\beta}{\sigma_{2}^{4}}}}{m_{TA_{2}}}}{\frac{V(TA_{1})\sqrt{1 + \frac{l'(\sigma_{1})^{2}}{4}+ TA_{1}^{2}l_{1}^{2}}\sigma_{1}^{2}\sqrt{1 + \frac{\beta}{\sigma_{1}^{4}}}}{m_{TA_{1}}}}e^{\frac{V(TA_{1})\sqrt{1 + \frac{l'(\sigma_{1})^{2}}{4}+ TA_{1}^{2}l_{1}^{2}}\sigma_{1}^{2}\sqrt{1 + \frac{\beta}{\sigma_{1}^{4}}}}{m_{TA_{1}}}\sigma_{1}} \cosh(\frac{V(TA_{1})\sqrt{1 + \frac{l'(\sigma_{1})^{2}}{4}+ TA_{1}^{2}l_{1}^{2}}\sigma_{1}^{2}\sqrt{1 + \frac{\beta}{\sigma_{1}^{4}}}}{m_{TA_{1}}}\tau_{1})} \times \nonumber\\&& \sinh(\frac{\frac{V(TA_{2})\sqrt{1 + \frac{l'(\sigma_{2})^{2}}{4}+ TA_{2}^{2}l_{2}^{2}}\sigma_{2}^{2}\sqrt{1 + \frac{\beta}{\sigma_{2}^{4}}}}{m_{TA_{2}}}}{\frac{V(TA_{1})\sqrt{1 + \frac{l'(\sigma_{1})^{2}}{4}+ TA_{1}^{2}l_{1}^{2}}\sigma_{1}^{2}\sqrt{1 + \frac{\beta}{\sigma_{1}^{4}}}}{m_{TA_{1}}}}e^{\frac{V(TA_{1})\sqrt{1 + \frac{l'(\sigma_{1})^{2}}{4}+ TA_{1}^{2}l_{1}^{2}}\sigma_{1}^{2}\sqrt{1 + \frac{\beta}{\sigma_{1}^{4}}}}{m_{TA_{1}}}\sigma_{1}} \times \nonumber\\&&\sinh(\frac{V(TA_{1})\sqrt{1 + \frac{l'(\sigma_{1})^{2}}{4}+ TA_{1}^{2}l_{1}^{2}}\sigma_{1}^{2}\sqrt{1 + \frac{\beta}{\sigma_{1}^{4}}}}{m_{TA_{1}}}\tau_{1})))\Big)\times\nonumber\\&& [4 \cosh^{2}\alpha_{III-A} + 5]\times \nonumber\\&&\cosh^{-4}\alpha_{III-A} ]^{2}
\label{a34}
\end{eqnarray}

\begin{eqnarray}
&& \rho_{IV-A}=\frac{3}{8\pi L_{P}^{2}} H_{IV-A}^{2} = \nonumber\\&& \frac{3}{8\pi L_{P}^{2}}[\sqrt{\frac{L_{P}^{2}}{4 \pi}}[\frac{1}{2}[\frac{4 T_{D3}^{2}}{\pi T_{0,IV-A}^{4}} \times \nonumber\\&&  [\frac{V(TA_{2})\sqrt{1 + \frac{l'(\sigma_{2})^{2}}{4}+ TA_{2}^{2}l_{2}^{2}}\sigma_{2}^{2}\sqrt{1 + \frac{\beta}{\sigma_{2}^{4}}}}{m_{TA_{2}}}]^{-2} \times \nonumber\\&& exp[4[E(TA_{1},TA_{2},\sigma_{1},\sigma_{2})]]\times \nonumber\\&&\Big( \cosh(- e^{-\frac{\frac{V(TA_{2})\sqrt{1 + \frac{l'(\sigma_{2})^{2}}{4}+ TA_{2}^{2}l_{2}^{2}}\sigma_{2}^{2}\sqrt{1 + \frac{\beta}{\sigma_{2}^{4}}}}{m_{TA_{2}}}}{\frac{V(TA_{1})\sqrt{1 + \frac{l'(\sigma_{1})^{2}}{4}+ TA_{1}^{2}l_{1}^{2}}\sigma_{1}^{2}\sqrt{1 + \frac{\beta}{\sigma_{1}^{4}}}}{m_{TA_{1}}}}e^{-a_{1}\sigma_{1}} \cosh(a_{1}\tau_{1})} \times \nonumber\\&&\sinh(-\frac{\frac{V(TA_{2})\sqrt{1 + \frac{l'(\sigma_{2})^{2}}{4}+ TA_{2}^{2}l_{2}^{2}}\sigma_{2}^{2}\sqrt{1 + \frac{\beta}{\sigma_{2}^{4}}}}{m_{TA_{2}}}}{\frac{V(TA_{1})\sqrt{1 + \frac{l'(\sigma_{1})^{2}}{4}+ TA_{1}^{2}l_{1}^{2}}\sigma_{1}^{2}\sqrt{1 + \frac{\beta}{\sigma_{1}^{4}}}}{m_{TA_{1}}}}e^{-\frac{V(TA_{1})\sqrt{1 + \frac{l'(\sigma_{1})^{2}}{4}+ TA_{1}^{2}l_{1}^{2}}\sigma_{1}^{2}\sqrt{1 + \frac{\beta}{\sigma_{1}^{4}}}}{m_{TA_{1}}}\sigma_{1}}\times \nonumber\\&& \sinh(\frac{V(TA_{1})\sqrt{1 + \frac{l'(\sigma_{1})^{2}}{4}+ TA_{1}^{2}l_{1}^{2}}\sigma_{1}^{2}\sqrt{1 + \frac{\beta}{\sigma_{1}^{4}}}}{m_{TA_{1}}}\tau_{1})))\Big)^{2}\times\nonumber\\&&\Big(\sinh^{2}(- e^{-\frac{\frac{V(TA_{2})\sqrt{1 + \frac{l'(\sigma_{2})^{2}}{4}+ TA_{2}^{2}l_{2}^{2}}\sigma_{2}^{2}\sqrt{1 + \frac{\beta}{\sigma_{2}^{4}}}}{m_{TA_{2}}}}{\frac{V(TA_{1})\sqrt{1 + \frac{l'(\sigma_{1})^{2}}{4}+ TA_{1}^{2}l_{1}^{2}}\sigma_{1}^{2}\sqrt{1 + \frac{\beta}{\sigma_{1}^{4}}}}{m_{TA_{1}}}}e^{-a_{1}\sigma_{1}} \cosh(a_{1}\tau_{1})} \times \nonumber\\&&\sinh(-\frac{\frac{V(TA_{2})\sqrt{1 + \frac{l'(\sigma_{2})^{2}}{4}+ TA_{2}^{2}l_{2}^{2}}\sigma_{2}^{2}\sqrt{1 + \frac{\beta}{\sigma_{2}^{4}}}}{m_{TA_{2}}}}{\frac{V(TA_{1})\sqrt{1 + \frac{l'(\sigma_{1})^{2}}{4}+ TA_{1}^{2}l_{1}^{2}}\sigma_{1}^{2}\sqrt{1 + \frac{\beta}{\sigma_{1}^{4}}}}{m_{TA_{1}}}}e^{-\frac{V(TA_{1})\sqrt{1 + \frac{l'(\sigma_{1})^{2}}{4}+ TA_{1}^{2}l_{1}^{2}}\sigma_{1}^{2}\sqrt{1 + \frac{\beta}{\sigma_{1}^{4}}}}{m_{TA_{1}}}\sigma_{1}}\times \nonumber\\&& \sinh(\frac{V(TA_{1})\sqrt{1 + \frac{l'(\sigma_{1})^{2}}{4}+ TA_{1}^{2}l_{1}^{2}}\sigma_{1}^{2}\sqrt{1 + \frac{\beta}{\sigma_{1}^{4}}}}{m_{TA_{1}}}\tau_{1})))+ \nonumber\\&& cosh^{2}(- e^{-\frac{\frac{V(TA_{2})\sqrt{1 + \frac{l'(\sigma_{2})^{2}}{4}+ TA_{2}^{2}l_{2}^{2}}\sigma_{2}^{2}\sqrt{1 + \frac{\beta}{\sigma_{2}^{4}}}}{m_{TA_{2}}}}{\frac{V(TA_{1})\sqrt{1 + \frac{l'(\sigma_{1})^{2}}{4}+ TA_{1}^{2}l_{1}^{2}}\sigma_{1}^{2}\sqrt{1 + \frac{\beta}{\sigma_{1}^{4}}}}{m_{TA_{1}}}}e^{-a_{1}\sigma_{1}} \cosh(a_{1}\tau_{1})} \times \nonumber\\&&\sinh(-\frac{\frac{V(TA_{2})\sqrt{1 + \frac{l'(\sigma_{2})^{2}}{4}+ TA_{2}^{2}l_{2}^{2}}\sigma_{2}^{2}\sqrt{1 + \frac{\beta}{\sigma_{2}^{4}}}}{m_{TA_{2}}}}{\frac{V(TA_{1})\sqrt{1 + \frac{l'(\sigma_{1})^{2}}{4}+ TA_{1}^{2}l_{1}^{2}}\sigma_{1}^{2}\sqrt{1 + \frac{\beta}{\sigma_{1}^{4}}}}{m_{TA_{1}}}}e^{-\frac{V(TA_{1})\sqrt{1 + \frac{l'(\sigma_{1})^{2}}{4}+ TA_{1}^{2}l_{1}^{2}}\sigma_{1}^{2}\sqrt{1 + \frac{\beta}{\sigma_{1}^{4}}}}{m_{TA_{1}}}\sigma_{1}}\times \nonumber\\&& \sinh(\frac{V(TA_{1})\sqrt{1 + \frac{l'(\sigma_{1})^{2}}{4}+ TA_{1}^{2}l_{1}^{2}}\sigma_{1}^{2}\sqrt{1 + \frac{\beta}{\sigma_{1}^{4}}}}{m_{TA_{1}}}\tau_{1})))\Big)\times\nonumber\\&& [4 \cosh^{2}\alpha_{IV-A} + 5]\times \nonumber\\&&\cosh^{-4}\alpha_{IV-A}  ]^{2}
\label{a35}
\end{eqnarray}

The above energy densities depend on the tachyonic potentials, the brane coordinates, the separation  between the branes and their accelerations. It is clear that when the energy of a universe in one region increases, the energy of the other universe in the other region decreases. This means that energy is exchanged between the two universes which causes the oscillation of the universes between two expanding and contracting phases (See figure 3).

\begin{figure*}[thbp]
	\begin{center}
		\begin{tabular}{rl}
			\includegraphics[width=8cm]{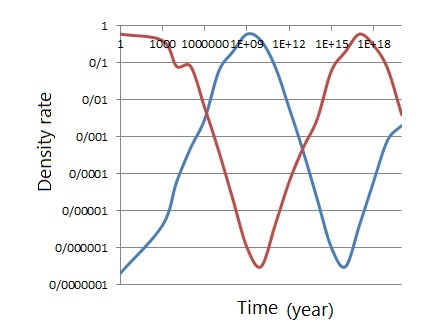}
		\end{tabular}
	\end{center}
	\caption{Density rate for universe in region III (blue color) and universe in region IV (red color) }
\end{figure*}

\section{Summary and Discussion} \label{sum}
Previously, we have shown that tachyonic fields cause the emergence of tachyonic BIons. A BIon is a system of two branes which are connected by a wormhole. These BIons accelerate and produce two regions. In this research, we have studied a system with two interacting BIons. We have shown that the interactions between the BIonic wormholes cause  the emergence of two new independent regions. The cosmic parameters like the Hubble parameter and energy density in each region depend on the parameters of the BIonic wormholes and act in opposite order to the cosmic parameters in the other region.

\acknowledgements{This work is based on the research supported wholly / in part by the National Research Foundation of South Africa (Grant Numbers: 118511)}

\end{document}